\documentclass[12pt]{article}
\usepackage[margin=1.25in]{geometry}
\usepackage{setspace}
\usepackage{amsmath,amssymb,amsthm,mathtools}
\usepackage{microtype}
\usepackage{tikz}
\usepackage{xcolor}
\usetikzlibrary{positioning,calc}
\usepackage{wasysym}
\usepackage[round]{natbib}
\usepackage[colorlinks=true,linkcolor=red,citecolor=blue,urlcolor=blue]{hyperref}
\newcommand{\blue}[1]{\textcolor{blue}{#1}}
\newcommand{\red}[1]{\textcolor{red}{#1}}
\theoremstyle{definition}
\newtheorem{definition}{Definition}
\newtheorem{theorem}{Theorem}

\newtheorem{lemma}[theorem]{Lemma}

\title{When are complements and substitutes dual?\thanks{I thank Christian Keeling, David Levine, and Klaus Ritzberger for useful discussions, and the Leverhulme Trust for financial support.}}
\author{Philip R. Neary\footnote{Royal Holloway, University of London. Email: \url{philip.neary@rhul.ac.uk}.}}
\date{\today}
\begin{document}
\maketitle
\begin{abstract}
\noindent
For binary-action games on bipartite networks, strategic complements and strategic substitutes are one and the same:~each can be transformed into the other by relabelling the actions of every player on exactly one side.
In this note, I show that bipartiteness is necessary and sufficient for such a relabelling; as a consequence, bipartite networks are precisely those on which every binary-action game with strategic substitutes must have a pure-strategy Nash equilibrium.
\end{abstract}


\section{Introduction}

Games in which actions are \emph{strategic complements} and games in which actions are \emph{strategic substitutes} are typically, and quite rightly, treated as distinct.
The terminology dates to \citet{BulowGeanakoplos:1985:JPE}, who distinguished between situations in which players' actions reinforce one another's incentives and those in which they offset.
Well-known examples appear in models of oligopoly, where prices are strategic complements in Bertrand competition while quantities are strategic substitutes in Cournot competition.

The distinction between the two classes also appears in their mathematical properties.
Games of strategic complements are particularly well-behaved:~the set of pure-strategy equilibria forms a nonempty lattice, and comparative statics are monotone \citep{Vives:1990:JME,MilgromShannon:1994:E,Topkis:1998:PUP}.
Games with strategic substitutes, on the other hand, do not generally enjoy the same guarantees.\footnote{The literature on strategic substitutes is less developed. Notable exceptions include \citet{RoySabarwal:2010:JME} and the monograph \citet{Sabarwal:2021:Palgrave}.}
In network games, however, the distinction between complements and substitutes may be less fundamental than it first appears.

In this note I consider when the distinction between the two classes is merely one of how players' actions are labelled.
To see that it can be, take a $2\times2$ game of either kind and swap the labels of one player's actions.
Hey presto!
No matter which type of game you began with, you have converted it into the other.
With three or more players, however, swapping the labels of one player's actions must work across every edge incident to that player, and this creates a consistency requirement that is, it turns out, not always satisfiable.
In other words: which, if any, network structures admit a globally consistent relabelling?

Theorem~\ref{thm:duality} gives a precise answer: bipartite networks, i.e., those in which players can be divided into two groups with all links between players of differing groups.\footnote{Bipartite networks appear throughout economics, for example to represent relationships between buyers and sellers, workers and firms, or firms and markets.}
If the network is bipartite, relabelling the actions of every player on one side means that every connection has exactly one relabelled endpoint, which reverses the strategic effect across that edge.
Conversely, any player-specific relabelling that transforms one class into the other must leave exactly one endpoint of every edge relabelled, a requirement that can be satisfied if and only if the graph contains no odd cycle.
Theorem~\ref{thm:existence} then shows that bipartite graphs are precisely those on which every binary-action game with strategic substitutes has a pure-strategy Nash equilibrium.\footnote{Related work by \citet{VanelliArditti:2025:arXiv} studies action relabelling in binary-action polymatrix games \citep{Yanovskaya:1968:LMS} on signed networks. The results here allow general payoff functions.}

For ease of exposition, I consider only binary-action games. 
But this restriction is not essential.
As long as action sets are totally ordered, reversing the order of actions of those on one side of a bipartite network transforms a game of strategic complements into one of strategic substitutes, and vice versa.

\section{Strategic complements and substitutes}

Let $G=(V,E)$ be a finite connected undirected graph.\footnote{All results extend componentwise to disconnected graphs.}
Each player $i\in V$ chooses an action $x_i\in\{a,b\}$, with $a$ understood as the ``lower'' action and $b$ as the ``higher'' one.
Player $i$'s utility from choosing $x_i$ is $u_i(x_i,\boldsymbol{x}_{N_i})$, where $N_i$ is the set of neighbours of $i$ in $G$ and $\boldsymbol{x}_{N_i}$ denotes their actions.
Utility functions are general in that I do not impose any pairwise-additive or separability assumption.

Strategic complements and strategic substitutes are often expressed through {\it increasing} and {\it decreasing differences}.
With only two actions, these conditions take a particularly simple form. For each player $i$, define
the function $D_i:\{a,b\}^{N_i}\to\mathbb{R}$ by
\[
D_i(\boldsymbol{x}_{N_i})= u_i(b,\boldsymbol{x}_{N_i})-u_i(a,\boldsymbol{x}_{N_i}).
\]
Now, for each neighbour $j\in N_i$, define the directed cross-effect function $\Delta_{ij}:\{a,b\}^{N_i\setminus\{j\}}\to\mathbb{R}$ by
\[
\Delta_{ij}(\boldsymbol{x}_{N_i\setminus\{j\}}) = D_i\bigl(x_j=b,\boldsymbol{x}_{N_i\setminus\{j\}}\bigr) - D_i\bigl(x_j=a,\boldsymbol{x}_{N_i\setminus\{j\}}\bigr).
\]
In words, $\Delta_{ij}$ measures how changing $j$'s action from $a$ to $b$, holding the actions of $i$'s other neighbours fixed, affects $i$'s incentive to choose action $b$.
Just because $G$ is undirected does not mean that the strategic effect of player $j$'s action on player $i$'s incentives are equal: $\Delta_{ij}$ and $\Delta_{ji}$ are different objects.
As regards the network structure, I assume that $ij\in E$ if and only if the two cross-effects $\Delta_{ij}$ and $\Delta_{ji}$ are not both everywhere zero.

\begin{definition}
A network game has \emph{strategic complements} if $\Delta_{ij}\geq 0$ everywhere, and \emph{strategic substitutes} if $\Delta_{ij}\leq 0$ everywhere.
\end{definition}

Now I formalise how to ``move between'' the two classes of game.
A \emph{relabelling} is a vector $r\in\{0,1\}^V$, where $r_i=0$
leaves player $i$'s action labels unchanged and $r_i=1$ exchanges
$a$ and $b$. Given a relabelling $r$, define
$T_r:\{a,b\}^V\to\{a,b\}^V$ as follows
\[
T_r(\boldsymbol{x})_i=
\begin{cases}
x_i, & r_i=0,\\
b, & r_i=1\text{ and }x_i=a,\\
a, & r_i=1\text{ and }x_i=b.
\end{cases}
\]
The relabelled game has payoff functions $u_i^{r}(\boldsymbol{x})=u_i\big(T_r(\boldsymbol{x})\big)$.
Note that applying the same relabelling twice always restores the original action profile, so $T_r$ is its own inverse: $T_r^{-1}=T_r$.

\section{When are complements and substitutes dual?}

The question boils down to when a globally consistent relabelling is possible.
To build up to that, I first consider the effect of relabelling actions along a single edge.

\begin{lemma}\label{lem:sign}
Under a player-specific action relabelling $r$,
\[
\Delta^r_{ij}(\boldsymbol{x}_{N_i\setminus\{j\}})
=
(-1)^{r_i+r_j}
\Delta_{ij}\bigl(T_r(\boldsymbol{x})_{N_i\setminus\{j\}}\bigr).
\]
In particular, the sign of the directed cross-effect is reversed if and only if exactly one of $i$ and $j$ is relabelled.
\end{lemma}

While Lemma~\ref{lem:sign} is almost immediate, it points to the real issue. Transforming strategic complements into strategic substitutes,
or vice versa, requires the actions of one, and only one, player on
every edge to be relabelled.
So when can local requirements of this nature can be applied consistently across a network?
The following provides the answer.

\begin{theorem}[Complements--substitutes duality]\label{thm:duality}
A network game with strategic complements can be transformed into one with strategic substitutes by a player-specific action relabelling, and vice versa, if and only if $G$ is bipartite.
When $G$ is bipartite, it suffices to relabel the actions of every player in one partite set.
\end{theorem}

The intuition behind the result is as follows.
Suppose first that $G$ is bipartite.
Relabelling the actions of every player on one side of the bipartition means that every edge has exactly one relabelled endpoint, so Lemma~\ref{lem:sign} applies on an ``edge by edge basis'' and every directed cross-effect is reversed.
The other direction is equally simple. Every edge must have one relabelled and one unrelabelled endpoint.
This can only be achieved consistently across a graph when it contains no odd cycle or, equivalently, when it is bipartite.
Since this is purely a relabelling of actions, best responses and equilibria correspond one-to-one.\footnote{I suspect that learning and adjustment results for strategic complements have analogs for substitutes on bipartite networks; see, for example, \citet{MilgromRoberts:1990:E}.}

\section{When is equilibrium guaranteed?}

Games with strategic complements always have at least one pure-strategy equilibrium whereas games with strategic substitutes need not.
I now show how Theorem~\ref{thm:duality} can be leveraged to give a graph-theoretic characterisation of when the complements guarantee extends to binary-action games with strategic substitutes.
It turns out that the answer is, once again, bipartite graphs.

\begin{theorem}\label{thm:existence}
A graph $G$ is bipartite if and only if every binary-action network game with strategic substitutes on $G$ has a pure-strategy Nash equilibrium.
\end{theorem}

One direction follows almost immediately from Theorem~\ref{thm:duality}.
Suppose $G$ is bipartite and fix a binary-action game with strategic substitutes.
Relabelling the actions of every player in one of the two partite sets transforms the game into one of strategic complements, which must have a pure-strategy Nash equilibrium \citep{Vives:1990:JME}.
Now undo the relabelling and you have a pure-strategy equilibrium of the original game.

The reverse is constructive.
Suppose $G$ is not bipartite and as such contains an odd cycle.
As it happens, it is always possible to engineer a game of strategic substitutes on $G$ such that any pure-strategy Nash equilibrium would require neighbouring players around the cycle to choose opposite actions, which is clearly impossible on an odd cycle.
While the problem generated by odd cycles is immediate, the issue of what to do with players off the cycle remains.
The trick is to make the remaining cross-effects of players on the cycle negative and sufficiently small, while giving players outside the cycle a strictly dominant action.
Thus, for every non-bipartite graph, one can always construct a network game with strategic substitutes that has no pure-strategy equilibrium.\footnote{Indeed, the counterexample can be constructed with every directed cross-effect strictly negative, so nonexistence does not rely on allowing zero cross-effects.}

The problem caused by odd cycles is illustrated below on a cycle of length five.
Vertices coloured \blue{blue} represent players that choose $a$, while those
coloured \red{red} choose $b$.
The requirement that neighbouring players choose opposite actions can be seen to break down on the edge $\{v_1,v_5\}$ that is marked in bold for emphasis.

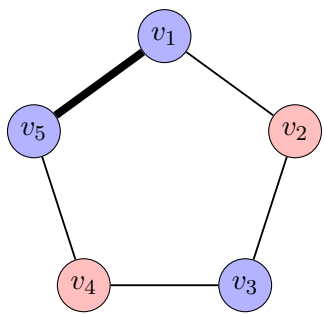
\begin{figure}[ht]
\centering
\begin{tikzpicture}[
    scale=0.7,
    vertex/.style={
        circle,
        draw=black,
        minimum size=7mm,
        inner sep=0pt,
        font=\small
    },
    actiona/.style={vertex, fill=blue!30},
    actionb/.style={vertex, fill=red!25},
    edge/.style={draw=black, line width=0.7pt},
    badedge/.style={draw=black, line width=3.0pt}
]

\node[actiona] (v1) at (90:2.6)  {$v_1$};
\node[actionb] (v2) at (18:2.6)  {$v_2$};
\node[actiona] (v3) at (-54:2.6) {$v_3$};
\node[actionb] (v4) at (-126:2.6) {$v_4$};
\node[actiona] (v5) at (162:2.6) {$v_5$};

\draw[edge]    (v1) -- (v2);
\draw[edge]    (v2) -- (v3);
\draw[edge]    (v3) -- (v4);
\draw[edge]    (v4) -- (v5);

\draw[badedge] (v5) -- (v1);



%

\end{tikzpicture}
\end{figure}

Relabelling actions does not create or destroy equilibria; all it does is give them different names.
Thus, on bipartite networks, games of strategic complements and strategic substitutes have exactly the same number of equilibria. The duality may then have more to offer than equilibrium existence: methods for finding all equilibria in games of strategic complements, such as those developed by \citet{Echenique:2007:JET}, might be put to work in games of strategic substitutes.

\appendix
\section{Proofs}\label{app:proofs}

\begin{proof}[Proof of Lemma~\ref{lem:sign}]
Fix a player $i\in V$ and a neighbour $j\in N_i$.
Holding the actions of all other neighbours fixed, recall that
\[
\Delta_{ij}
=
\bigl[u_i(b,b)-u_i(a,b)\bigr]
-
\bigl[u_i(b,a)-u_i(a,a)\bigr],
\]
where the first argument is $i$'s action and the second is $j$'s.

Relabelling player $i$'s actions reverses the order of the
payoffs within each bracket, while relabelling player $j$'s
actions exchanges the two brackets.
Thus either operation multiplies $\Delta_{ij}$ by $-1$,
whereas doing both leaves it unchanged.
Relabelling any other neighbour simply changes the profile at which $\Delta_{ij}$ is evaluated, completing the proof.
\end{proof}

\begin{proof}[Proof of Theorem~\ref{thm:duality}]
I show how to relabel a network game of strategic substitutes into one with strategic complements; the other direction is analogous.

First, suppose $G=(L\cup R,E)$ is bipartite and consider a game with strategic substitutes. Choose a relabelling $r$ by setting $r_i=0$ for $i\in L$ and $r_i=1$ for $i\in R$. Every edge then has exactly one relabelled endpoint, so Lemma~\ref{lem:sign} implies that every directed cross-effect is reversed, and it follows that actions in the relabelled game are strategic complements.

Conversely, suppose there exists a player-specific relabelling $r$ that transforms a game with strategic substitutes into one with strategic complements. For every $ij\in E$, at least one of $\Delta_{ij}$ and $\Delta_{ji}$ is not identically zero and must therefore be reversed. Lemma~\ref{lem:sign} then implies that $r_i\neq r_j$ for every $ij\in E$.
It follows that the relabelled and unrelabelled players form a bipartition of $G$.
\end{proof}

\begin{proof}[Proof of Theorem~\ref{thm:existence}]
Suppose first that $G$ is bipartite. By Theorem~\ref{thm:duality},
any binary-action game with strategic substitutes on $G$ can be
transformed by relabelling the actions of every player in one partite
set into a game with strategic complements. Such a game has a
pure-strategy Nash equilibrium \citep{Vives:1990:JME} and undoing the
relabelling gives a pure-strategy equilibrium of the original
game.

For the converse, suppose that $G$ is not bipartite.
Then $G$ must contain an odd cycle $C=(v_1,v_2,\ldots,v_{2k+1},v_1)$.
I construct a binary-action game with strategic substitutes on $G$ that has no pure-strategy Nash equilibrium.
The construction treats players differently according to whether or not they lie on $C$.

For each $v_i\in C$, set
\[
u_{v_i}(a,\boldsymbol{x}_{N_{v_i}})=0
\]
and
\[
u_{v_i}(b,\boldsymbol{x}_{N_{v_i}})
=
d_i-\frac12
-d_i\mathbf{1}\{x_{v_{i-1}}=b\}
-\sum_{j\in N_{v_i}\setminus\{v_{i-1}\}}
\mathbf{1}\{x_j=b\},
\]
where $v_0=v_{2k+1}$.
It is easily verified that, irrespective of the actions of her other neighbours, player $v_i$'s unique best response is always to anti-coordinate with $v_{i-1}$.

Now, for each player $i$ not on the cycle, set
\[
u_i(a,\boldsymbol{x}_{N_i})=0,
\qquad \text{and} \qquad
u_i(b,\boldsymbol{x}_{N_i})
=
-1-\sum_{j\in N_i}\mathbf{1}\{x_j=b\}.
\]
Since $u_i(b,\boldsymbol{x}_{N_i})<0$ everywhere, action $a$ is strictly dominant for these players.

In sum, all directed cross-effects in the constructed game are negative and so actions are strategic substitutes.
Moreover, in any equilibrium, players on $C$ would have to alternate between $a$ and $b$, which is impossible on an odd cycle.
Hence the constructed game has no pure-strategy Nash equilibrium, and so we are done.
\end{proof}


\bibliographystyle{plainnat}
\bibliography{dual.bib}
\end{document}